\documentclass[twocolumn,showpacs,preprintnumbers,amsmath,amssymb,floatfix]{revtex4}
\input{epsf}

\usepackage{graphicx}
\usepackage{dcolumn}
\usepackage{bm}

\begin{document}

\title{Entangled light from Bose-Einstein condensates}
\author{H. T. Ng and S. Bose}
\affiliation{{Department of Physics and Astronomy, University
College London, Gower Street, London WC1E 6BT, United Kingdom}}
\date{\today}

\begin{abstract}
We propose a method to generate entangled light with a Bose-Einstein
condensate trapped in a cavity, a system realized in recent
experiments. The atoms of the condensate are trapped in a periodic
potential generated by a cavity mode. The condensate is continuously
pumped by a laser and spontaneously emits a pair of photons of
different frequencies in two distinct cavity modes. In this way, the
condensate mediates entanglement between two cavity modes which leak
out and can be separated and exhibit continuous variable entanglement. 
The scheme exploits the experimentally demonstrated strong, steady 
and collective coupling of condensate atoms to a cavity field.
\end{abstract}


\maketitle

Quantum communications can outperform their classical counterparts,
for example, quantum cryptography enables secure distribution of
quantum information. ``Quantum correlations" or entanglement, when
shared between distant parties, is a key resource for quantum
communication tasks such as quantum cryptography \cite{Ekert},
teleportation \cite{bennett93} and dense coding \cite{bennett92}.
These applications provide a very strong motivation for entangled
light which is widely regarded as the most ideal entity for the
sharing of entanglement between genuinely distant parties
\cite{Bouwmeester}. Additionally, as the interface between light and
matter matures as a technology \cite{duan,polzik}, entangled light
can also link up distinct matter registers of a quantum computer and
thereby aid in scaling up quantum computers.

One important form of entangled light \cite{Reid,Schori} is continuous variable (CV) entanglement between phase
quadratures of two distinct modes of the light field of the type
discussed in the famous Einstein-Podolsky-Rosen (EPR) paper \cite{EPR}.
Such entanglement has been used for quantum teleportation and has applications 
in quantum dense coding \cite{ban-dense} and
quantum cryptography \cite{ralph}. Additionally, if the entanglement is sufficiently 
``narrow band" in frequencies then the quantum states of the light will efficiently interface with
those of atomic ensembles \cite{polzik} for applications in quantum repeaters and linking 
quantum registers. Thus the motivation for having entangled sources of EPR light is very strong.

The prevalent sources of EPR entangled light are non-linear crystals. 
It was noticed long ago that light fields
produced from nonlinear crystals seem to be non-classically
correlated \cite{Burnham}. For non-degenerate optical parametric
oscillators, Reid demonstrated that the quadrature phases of the
output fields have EPR type entanglement \cite{Reid} and this is
indeed one of the sources in recent experiments \cite{Schori}.
Alternatively, the outputs of a two degenerate optical parametric
oscillators are interfered to obtain EPR entangled light
\cite{kimble}. For such crystals, the Hamiltonian is actually
phenomenologically constructed to describe the observed nonlinear
processes and expressed in terms of the nonlinear susceptibility of
the macroscopic crystal. This is why there has been a recent
interest in deriving EPR entangled light from a more fundamental
Hamiltonian, such as from the ``quantized motion" of a single atom
trapped in a cavity \cite{Morigi,Vitali}. This provides a
coherent control of the entanglement generator at the microscopic
system, as opposed to a bulk crystal. In addition to this
fundamental interest, such alternative sources may also have a
practical interest if they can surpass the squeezing parameter (a
parameter that controls the amount of entanglement in the EPR
entangled light) possible from crystal sources as many of the
restrictions such as the lower finesse of cavities around crystals,
or the unbalanced absorption of the entangled modes while traversing
the crystal, do not directly apply.

\begin{figure}[ht]
\includegraphics[height=4.5cm]{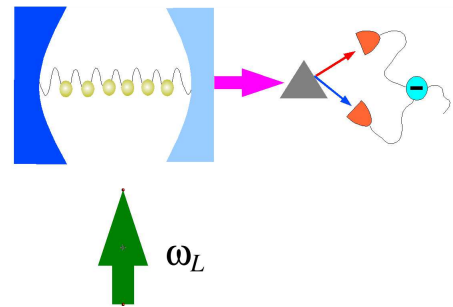}
\caption{\label{fig1} Configuration: A two-component condensate is
trapped inside a cavity. The condensate is pumped by the external
laser field of the frequency $\omega_L$. The cavity field is decayed
through the one-sided cavity.  The output cavity field are then
split into two separate modes by a prism and probed by the homodyne
detection.}
\end{figure}

Recently, the strong coupling of an atomic Bose-Einstein condensate
(BEC) to a single-mode photon field of an optical cavity has been
experimentally achieved \cite{Brennecke,Colombe}.  The ultracold
atoms are trapped in a periodic potential generated by a quantized
field mode \cite{Maschler}. Since the $N$ two-level atoms are
identically coupled to the single-mode photon field which gives a
collective enhancement of a factor $\sqrt{N}$
\cite{Brennecke,Colombe}. In fact, such strong atom-photon couplings
are very useful in performing quantum information processing (QIP)
before the decoherence sets in.  The potential applications include
long-lived quantum memory \cite{Schori} and quantum network for
light-matter interface \cite{Duan0}.

In this paper, we consider an atomic BEC trapped inside an optical cavity \cite{Brennecke}.
Each atom is located at the anti-node of the quantized cavity field so that one atom per site can identically couple
to the cavity field.  Compared to a thermal cloud of atoms with the inhomogeneous atom-photon couplings
in the cavity, we can truely apply the Tavis-Cummings model \cite{Tavis} to study our system.
In addition, the BEC's with reduced Doppler broadening leads to much longer
coherence times than that of the thermal clouds \cite{Inouye}.

We consider the BEC is continuously driven by an external laser and spontaneously emits photons with the
two different frequencies in pair.  Hence, the BEC acts as a medium to mediate
the entanglement between the two cavity modes.  The two quantum-correlated light modes are emitted
through a one-sided mirror as shown in Fig. \ref{fig1}.   Since the ultracold atoms have long coherence times \cite{Harber},
this can provide a robust way in generating the entangled light.  We will show that the degree of
entanglement depends on the ratio of the decay rate of the cavity and the effective Rabi frequency.  This means that
the degree of entanglement can be controlled by adjusting the atom-photon coupling strength.

\begin{figure}[ht]
\includegraphics[height=3.5cm]{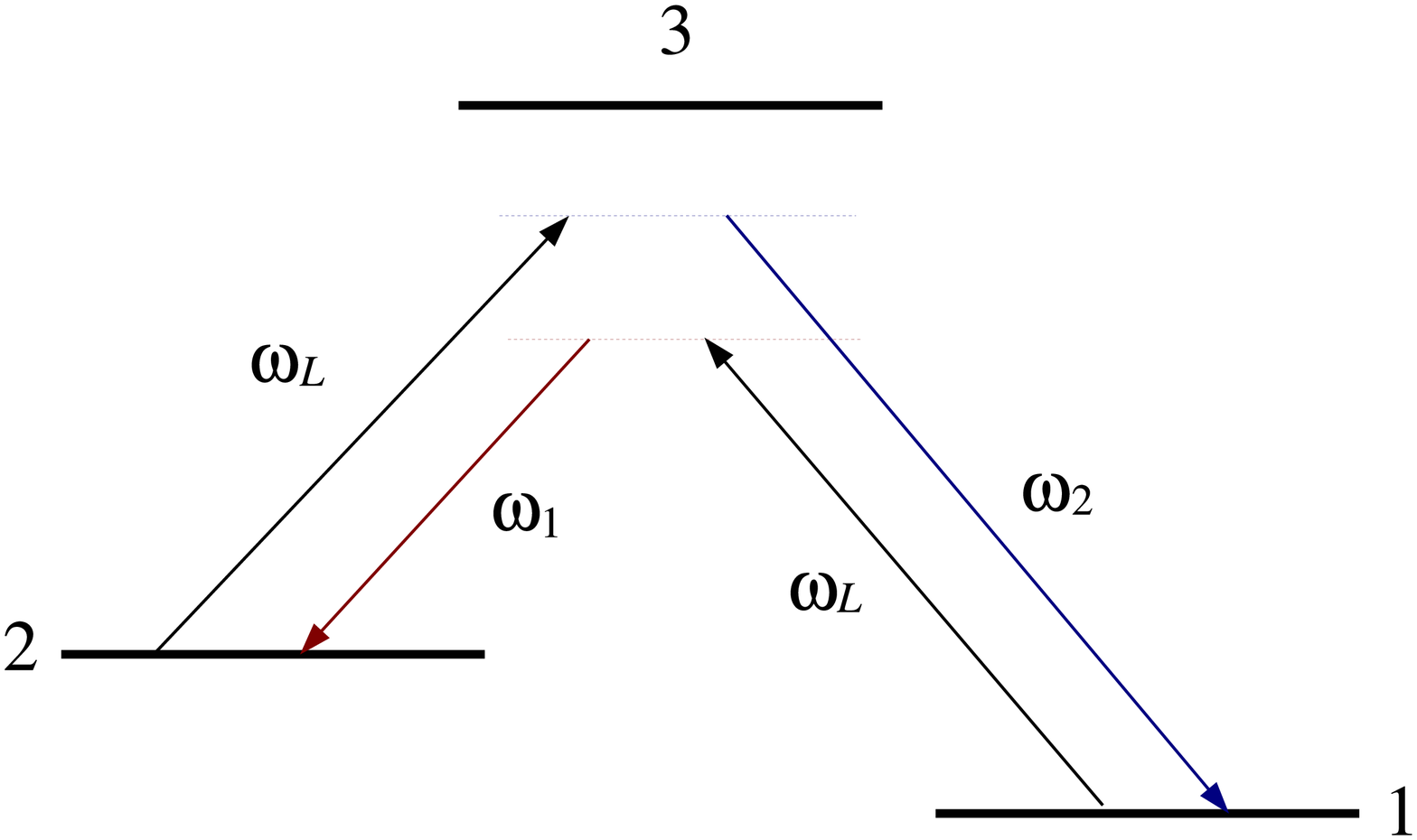}
\caption{ \label{fig2} Energy level of an atom:  A laser with frequency $\omega_L$
is applied to continuously pump the
levels $|1\rangle$ and $|2\rangle$ respectively.  The state $|1\rangle{(|2\rangle)}$ is coupled with the
state $|3\rangle$ and interact with the cavity field
$a_{1,2}$. }
\end{figure}

We consider a two-component condensate trapped inside a cavity in
which the atoms are trapped in a one-dimensional optical lattice as
shown in Fig. \ref{fig1}. A classical laser, with frequency
$\omega_L$, is used to pump the two internal states $|1\rangle$ and
$|2\rangle$ to a higher level $|3\rangle$. Then, the two different
quantized light fields with frequencies $\omega_1$ and $\omega_2$
are spontaneously emitted due to large detuning \cite{Law}.  The Hamiltonian is
written as
\begin{eqnarray}
H&=&\sum^{N}_{k=1}(\omega_{31}|3\rangle{}_{k}\langle{3}|+\omega_{21}|2\rangle{}_k\langle{2}|)+
\sum^{N}_{k=1}\sum^2_{j=1}[\omega_{j}a^\dag_{j}a_{j}\nonumber\\
&&+\Omega_j(|j\rangle_k\langle{3}|e^{i\omega_Lt}+{\rm
H.c.})+\lambda_{j}(a^\dag_j|j\rangle_k\langle{3}|+{\rm H.c.})],\nonumber\\
\end{eqnarray}
where $\omega_{j1}$  are the energy splitting between the states
$|j\rangle_k$ and $|1\rangle_k$.  The frequencies of the two modes
satisfy the two-photon Raman resonance condition
$2\omega_L=\omega_1+\omega_2$ \cite{Law} so that we can write
$\omega_{1,2}=\omega_L\pm{\nu}$.  These two modes must satisfy the
boundary condition of the cavity.

It is instructive to work in the rotating frame by using the unitary
transformation
$U(t)=\exp[i\omega_L(\sum^{N}_{k=1}|3\rangle_k\langle{3}|+\sum^{2}_{j=1}a^\dag_j{a}_j)t]$.
The transformed Hamiltonian reads as
\begin{eqnarray}
\tilde{H}&=&\sum^{N}_{k=1}(\Delta|3\rangle{}_{k}\langle{3}|+\omega_{21}|2\rangle{}_k\langle{2}|)+
\sum^{N}_{k=1}\sum^2_{j=1}[\delta_{j}a^\dag_{j}a_{j}\nonumber\\
&&+\Omega_j(|j\rangle_k\langle{3}|+{\rm
H.c.})+\lambda_{j}(a^\dag_j|j\rangle_k\langle{3}|+{\rm H.c.})],
\end{eqnarray}
where $\Delta=\omega_{31}-\omega_L$ and $\delta_j=\omega_j-\omega_L$.
For $\Delta\gg\Omega_j,\lambda_j$, this enables us to adiabatically
eliminate the upper level $|3\rangle$.

The effective Hamiltonian is given by
\begin{equation}
H'=\sum^2_{j=1}\delta_j{a}^\dag_ja_j+\tilde{\omega}J_z-[(g_1a_1+g_2a^\dag_2)J_++{\rm
H.c.}],
\end{equation}
where $J_{+}=\sum^{N}_{k=1}|2\rangle_{k}\langle{1}|$ and
$J_z=\sum^N_{k=1}(|2\rangle_k\langle{2}|-|1\rangle_k\langle{1}|)$.
The parameters $\tilde{\omega}$ and $g_j$ are
$\omega_{21}+(\Omega^2_1-\Omega^2_2)/\Delta$ and
$\lambda_j\Omega_j/\Delta$ respectively.  We note that
$C=J_z+a^\dag_1a_1-a_2a^\dag_2$ is a constant of motion.

We consider that the low-lying collective excitations in the condensates
involve throughout the dynamics.  Then, we can approximate the
angular momentum operator as a harmonic oscillator as \cite{Ng}:
$J_+{\approx}\sqrt{N}b^\dag$, $J_-{\approx}\sqrt{N}b$ and
$J_z{\approx}b^\dag{b}-N/2$.  The Hamiltonian can be rewritten as
\begin{eqnarray}
H'&\approx&\omega'b^\dag{b}-[\sqrt{N}(g_1a_1+g_2a^\dag_2)b^\dag+{\rm
H.c.}].
\end{eqnarray}
We assume $\omega'=(\tilde{\omega}-\nu)\gg{g_j}\sqrt{N}$ such that
the low excitations approximation is valid.   In the large detuning
limit, we can write
\begin{eqnarray}
H_{\rm
eff}&=&\chi_1a^\dag_1a_1+\chi_2a_2a^\dag_2+{\chi}(a^\dag_1a^\dag_2+{a}_1a_2),
\end{eqnarray}
where $\chi_j=g^2_jN/\omega'$ and
$\chi=g_1g_2{N}/\omega'$.

For convenience, we represent the Hamiltonian in terms of the
operators \cite{Barnett}:
\begin{eqnarray}
K_{3}&=&\frac{1}{2}(a^\dag_1a_2+a_2a^\dag_2), \\
K_+&=&a^\dag_1a^\dag_2=K^\dag_-.
\end{eqnarray}
These operators satisfy the commutation relations $[K_3,K_{\pm}]=\pm
K_{\pm}$ and $[K_+,K_-]=-2K_3$.   Thus, the Hamiltonian is
represented in the form as (we have ignored the constant term):
\begin{eqnarray}
H_{\rm eff}&=&\frac{\chi_1+\chi_2}{2}K_3+{\chi}(K_++K_-).
\end{eqnarray}

We consider the initial state is the vacuum state and thus the time
evolution of two-mode state is 
$|\Psi(\tau)\rangle=e^{-iH_{\rm eff}\tau}|0,0\rangle$.
According to the operator ordering theorem \cite{Barnett}, we have
\begin{eqnarray}
e^{-iH_{\rm
eff}\tau}&=&e^{{\Gamma}K_+}e^{\ln{\tilde{\Gamma}}K_3}e^{{\Gamma}K_-},
\end{eqnarray}
where
\begin{eqnarray}
\tilde{\Gamma}&=&\Big(\cosh\beta-\frac{\tilde{\gamma}}{2\beta}\sinh\beta\Big)^{-2},\\
{\Gamma}&=&\frac{2\gamma\sinh\beta}{2\beta\cosh\beta-\tilde{\gamma}\sinh\beta}.
\end{eqnarray}
The parameters $\gamma$, $\tilde{\gamma}$, $\beta^2$ are
$-i\chi\tau$, $-i({\chi_1+\chi_2})\tau/{2}$ and ${\tilde{\gamma}^2}/{4}-\gamma^2$ respectively.
Therefore, the state can be readily obtained as
\begin{eqnarray}
|\Psi(\tau)\rangle=\tilde{\Gamma}\sum^{\infty}_{n=0}\Gamma^n|n,n\rangle.
\end{eqnarray}
Clearly, this state is a entangled state in which the two cavity modes are
entangled in pair.

We have shown that the entanglement between the two cavity modes can be produced inside the cavity. However, it is necessary to
detect the entangled light out of the cavity.  The entangled light is emitted through the mirror and splits into the two different
frequency components via a prism as shown in Fig \ref{fig1}.
They are measured via a homodyne detection.  The difference in the photon current is then recorded.
The resulting squeezing spectrum can be obtained by
the spectrum analyser \cite{Walls}.

To evaluate the entangled light outside the cavity, we thus have to take account of the input-output theory \cite{Walls}.
The Langvein equations of motion for the system are given by \cite{Gardiner}
\begin{eqnarray}
\dot{a}_j&=&ig_j\sqrt{N}b-\kappa_ja_j-\sqrt{2\kappa_j}a_{j\rm in}, \\
\dot{b}&=&-i\omega'b+i\sqrt{N}(g_1a_1+g_2a^\dag_2).
\end{eqnarray}
The output fields are $a_{jo}=a_{j\rm{in}}+\sqrt{2\kappa_j}a_j$.  We assume that the radiative noise from the cavity is much larger than the noise
coming from the BEC and also the input noise source is in vacuum.

Now we study the squeezing spectrum by transforming to the Fourier space
as \cite{Vitali}:
\begin{eqnarray}
I^\theta_{\pm}(\omega)=\frac{1}{\sqrt{2\pi}}\int{dt}e^{-i\omega{t}}I^\theta_{\pm}(t),
\end{eqnarray}
where $I^\theta_+=a_{\rm 1o}e^{-i\theta}+a^\dag_{\rm 1o}e^{i\theta}-a_{\rm 2o}e^{-i\theta}-a^\dag_{\rm 2o}e^{i\theta}$ and
$I^\theta_-=-i[a_{\rm 1o}e^{-i\theta}-a^\dag_{\rm 1o}e^{i\theta}+a_{\rm 2o}e^{-i\theta}-a^\dag_{\rm 2o}e^{i\theta}]$.
The squeezing spectrum can be defined as
$\langle{{I}^{\theta}_{\pm}(\omega){I}^{\theta}_{\pm}(\omega')+{I}^{\theta}_{\pm}(\omega'){I}^{\theta}_{\pm}(\omega)}\rangle=2S^{\theta}_{\pm}\delta{(\omega+\omega')}$.
In our case, we found that $S^{\theta}_+(\omega)=S^{\theta}_-(\omega)$.
This can be shown that the two output modes are entangled if $S^\theta_{\pm}(\omega)<1$ \cite{Vitali,Duan}.

\begin{figure}
\centering
\includegraphics[height=6.5cm]{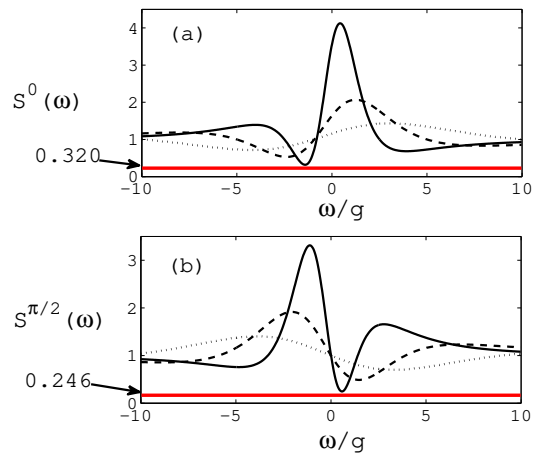}
\caption{\label{fig3} (Color online) The squeezing spectrums $S^0(\omega)$ and $S^{\pi/2}(\omega)$ are plotted as a function of the frequency $\omega/g$, where $g_1=g_2=g$,
$\omega=10^{4}g$ and $N=10^4$.  The solid, dash and dotted lines represent the different values of $\kappa=10,5,2.5$ (in units of $g$) respectively.}
\end{figure}

We investigate the squeezing spectrum $S^\theta(\omega)$  for the regime of $\omega'\gg{g_{1,2}\sqrt{N}}$.   For simplicity, we take $g_{1,2}=g$ and
$\kappa_{1,2}=\kappa$.  In Fig. \ref{fig3} (a), we plot the squeezing spectrum $S^0(\omega)$ as a function of $\omega$ (in units of $g$)
for the different decay rates $\kappa$.  It shows that the squeezing occurs at the negative frequency domain whereas the unsqueezing occurs at the positive
frequency domain.  We also plot the out-of-phase squeezing spectrum $S^{\pi/2}(\omega)$ in Fig. \ref{fig3} (b).  In contrast, the squeezing(unsqueezing) occurs
at the positive(negative) frequency domain.  Apart from that, we can see that a larger squeezing compensates a narrower range of frequencies as shown in
Fig. \ref{fig3} (a) and (b).

\begin{figure}
\centering
\includegraphics[height=5cm]{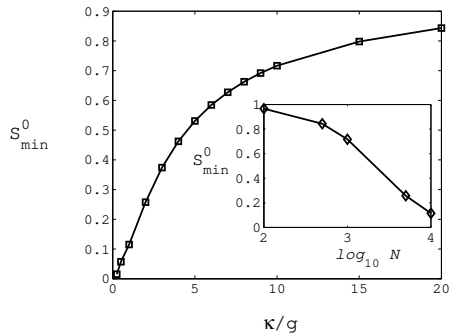}
\caption{\label{fig5}  The minimal values of squeezing for the squeezing spectrum $S^{0}_{\rm min}$ are plotted as a function of $\kappa/g$, where,
$\omega'=10^{4}g$ and $N=10^4$.  The inset shows the minimal squeezing $S^0_{\rm min}$ as a function of $log_{10}N$ for $\kappa=g$ and $\omega'=10^4g$.}
\end{figure}

We study the maximal degree of squeezing attainable in the system for the values of cavity decay and the number of atoms.
The minimal values of squeezing for the spectrum $S^0(\omega)$ as a function of $\kappa$ (in units of $g$) as shown in Fig. \ref{fig5}.
The degree of squeezing increases significantly as the cavity decay parameter $\kappa$ decreases.  The nearly perfect squeezing can be obtained for in the limit of
$\kappa$ approaching to zero.  In the inset of Fig. \ref{fig5}, the minimal values of squeezing is plotted as a function of $log_{10}N$.
This means that the strong squeezing can be attained as the number of atoms increases.

\begin{figure}
\centering
\includegraphics[height=4.5cm]{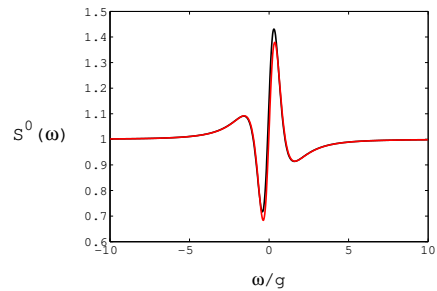}
\caption{\label{fig6} (Color online) The squeezing spectrum $S^0(\omega)$ are plotted as a function of the frequency $\omega/g$, where 
$\omega'=10^{5}g$ and $N=10^4$.  The black and red  lines represent the exact and approximated solutions for $\kappa=g$.}
\end{figure}

For $\omega'\gg{g^2_{1,2}{N}}$,  the approximated analytical expression of the squeezing spectrum $S(\omega)$ can
be found as
\begin{widetext}
\begin{eqnarray}
\label{appS}
S(\omega)&\approx&1+\frac{4\kappa^2g_1g_2N}{(\kappa^2+\omega^2)^2({\omega'}^2-\omega^2)}\Big[\frac{4\omega\omega'(\kappa^2-\omega^2)}{\kappa^2+\omega^2}+\frac{g_1g_2N({\omega'}^2+\omega^2)}{{\omega'}^2-\omega^2}\Big].
\end{eqnarray}
\end{widetext}
The approximate solution Eq.(\ref{appS}) is compared with the numerical solution in Fig \ref{fig6}.
This shows that it is a very good approximation in the limit of this large detuning.

We now briefly examine the range of parameters available for our
scheme if the setups of some recently performed experiments are
directly used. If one literally uses the parameters from the recent
Brennecke {\em et. al.} experiment \cite{Brennecke} then one can
have $\lambda_j=2\pi\times10.6\approx 67$ MHz (what we say below
holds for both $j=1,2$). Typically, $\Omega_j$, can be made even
larger as it is proportional to the strength of the external laser
field, so we assume it to be $\eta\lambda_j$, where $\eta$ is a
numerical factor which can be varied between $1$ and $4$. The
detuning $\Delta^2>> \lambda_j^2,\Omega_j^2,\lambda_j\Omega_j$ is
required for the adiabatic elimination of the level $|3\rangle$. So
we choose $\Delta \sim 10 \lambda_j$. Thus the effective Rabi
frequency $g_j\sim\lambda_j{\Omega_j}/\Delta=6.7\eta$ MHz. The
cavity decay $\kappa =2\pi\times 1.3\approx 8.1$MHz. Thus the ratio
$\kappa/g_j$ can be made to vary between $1$ and $0.3$ by varying
$\eta$ between $1$and $4$. The energy splitting $\omega'\sim
\Delta_{12}$ is around 10 GHz (which means $\omega'$, when expressed
in terms of $g_j$ is $10^3-10^4 g_j$). Though the number of atoms
$N$ can be up to $2\times 10^5$ \cite{Brennecke}, we restrict the
number to about $\sim 10^4$ which is also a number in typical
experiments, so that $\omega'>> g_j\sqrt{N}$ is fulfilled and the
BEC is not excited. We noted that the cavity decay rate $\kappa$ can
be adjusted because $\kappa=\pi{c}/(2LF)$ depends on the length $L$
and the finesse $F$ of the cavity \cite{Hinds}. Hence, the different
extent of squeezing can be observed in experiment by varying the
parameters $\kappa$ and $g$.

We have studied the entangled light generation with the cavity-BEC system in which
the atoms are continuously driven by the external laser field.   The entangled light can
emit from the cavity through the one-side mirror and then the entangled light can be measured via the homodyne detection.
We have shown that the degree of entanglement can be controlled by adjusting the
strength of atom-photon couplings.  Our scheme to generate entangled light can be realized with the current experimental
 technology \cite{Brennecke,Colombe}.

\begin{acknowledgments}
The work of H. T. Ng is supported by the Quantum Information
Processing IRC (QIPIRC) (GR/S82176/01). S. Bose also thanks the
Engineering and Physical Sciences Research Council (EPSRC) UK and
the Royal Society and the Wolfson Foundation.

\end{acknowledgments}

\end{document}